\documentclass[3p,,preprint,12pt]{elsarticle}
\usepackage{lineno,hyperref}

\usepackage{amsmath,color}
\usepackage{amssymb}
\usepackage{graphicx}
\usepackage{bm}
\usepackage{subcaption}
\usepackage{natbib}
\usepackage{booktabs}
\usepackage{longtable}
\usepackage{array}
\newcolumntype{x}[1]{%
>{\centering\hspace{0pt}}p{#1}}%
\usepackage{multirow}
\usepackage[table]{xcolor}
\usepackage{float}
\usepackage{colortbl}
\usepackage{pdflscape}
\usepackage{threeparttable}
\usepackage{verbatim}

\def\newpage{\vfill\eject}

\def\today{\ifcase\month\or
  January\or February\or March\or April\or May\or June\or
  July\or August\or September\or October\or November\or December\fi
  \space\number\day, \number\year}

\newdimen\biblioindent    \biblioindent=30pt

 at 7truept

\usepackage{mathtools}
\usepackage{amsthm}
\usepackage[mathscr]{eucal}
\usepackage{bm}
\usepackage{eqlist} % Makes for a nice list of symbols.
\usepackage{empheq}
\usepackage{algorithm}
\usepackage{algorithmic}
\usepackage{cases}
\usepackage{comment}
\usepackage{inputenc}
\usepackage{pifont}
\usepackage{hyperref}
\usepackage{amsmath}
\usepackage{amssymb}
\usepackage{amsthm}
\usepackage{exscale}
\usepackage[mathscr]{eucal}
\usepackage{bm}
\usepackage{eqlist} % Makes for a nice list of symbols.
\usepackage{empheq}
\usepackage{algorithm}
\usepackage{algorithmic}
\usepackage{cases}
\usepackage{indentfirst} 
\usepackage{comment}
\usepackage{pifont}
\usepackage{natbib}
\usepackage{hyperref}
\usepackage{lipsum}

\DeclareGraphicsExtensions{.pdf, .jpg}

\newcommand{\bA}{\bm{A}}

\newcommand{\bX}{\bm{X}}

\newcommand{\bx}{\bm{x}}

\newcommand{\by}{\bm{y}}

\newcommand{\bz}{\bm{z}}
\newcommand{\bomega}{\bm{\omega}}

\newcommand{\bSigma}{\bm{\Sigma}}

\newcommand{\bgamma}{\bm{\gamma}}
\newcommand{\mL}{\mathcal{L}}

\usepackage{enumitem}

\newcommand{\bbeta}{\bm{\beta}}

\newcommand{\balpha}{\bm{\alpha}}

\newcommand{\be}{\begin{equation}}
\newcommand{\ee}{\end{equation}}
\newcommand{\beq}{\begin{equation}}
\newcommand{\eeq}{\end{equation}}
\newcommand{\beqn}{\begin{eqnarray}}
\newcommand{\eeqn}{\end{eqnarray}}
\newcommand{\beqnn}{\begin{eqnarray*}}
\newcommand{\eeqnn}{\end{eqnarray*}}

\newcommand{\mA}{\mathcal{A}}

\newcommand{\mR}{\mathcal{R}}
\newcommand{\mT}{\mathcal{T}}

\DeclareMathOperator*{\argmin}{\arg\!\min}

\usepackage{scalerel,stackengine}
\stackMath
\newcommand\myhat[1]{%
\savestack{\tmpbox}{\stretchto{%
  \scaleto{%
    \scalerel*[\widthof{\ensuremath{#1}}]{\kern-.6pt\bigwedge\kern-.6pt}%
    {\rule[-\textheight/2]{1ex}{\textheight}}%WIDTH-LIMITED BIG WEDGE
  }{\textheight}% 
}{0.5ex}}%
\stackon[1pt]{#1}{\tmpbox}%
}

\newtheorem*{remark}{Remark}

\newtheorem{theorem}{Theorem}

\def\citeay#1{\citeauthor{#1} \cite{#1}}
\usepackage{url,multirow,morefloats,floatflt,cancel,tfrupee}
\makeatletter

\AtBeginDocument{\@ifpackageloaded{textcomp}{}{\usepackage{textcomp}}}
\makeatother
\usepackage{colortbl}
\usepackage{xcolor}
\usepackage{pifont}
\usepackage[nointegrals]{wasysym}
\makeatletter

\def\mcWidth#1{\csname TY@F#1\endcsname+\tabcolsep}

\def\cAlignHack{\rightskip\@flushglue\leftskip\@flushglue\parindent\z@\parfillskip\z@skip}
\def\rAlignHack{\rightskip\z@skip\leftskip\@flushglue \parindent\z@\parfillskip\z@skip}

\@ifundefined{etal}{}{}

\usepackage{ifxetex}
\ifxetex\else\if@twocolumn\@ifpackageloaded{stfloats}{}{\usepackage{dblfloatfix}}\fi\fi

\AtBeginDocument{
\expandafter\ifx\csname eqalign\endcsname\relax
\def\eqalign#1{\null\vcenter{\def\\{\cr}\openup\jot\m@th
  \ialign{\strut$\displaystyle{##}$\hfil&$\displaystyle{{}##}$\hfil
      \crcr#1\crcr}}\,}
\fi
}

\AtBeginDocument{%
  \@ifpackageloaded{endfloat}%
   {\renewcommand\efloat@iwrite[1]{\immediate\expandafter\protected@write\csname efloat@post#1\endcsname{}}}{\newif\ifefloat@tables}%
}%

\def\BreakURLText#1{\@tfor\brk@tempa:=#1\do{\brk@tempa\hskip0pt}}
\let\lt=<
\let\gt=>
\def\processVert{\ifmmode|\else\textbar\fi}

\@ifundefined{subparagraph}{
\def\subparagraph{\@startsection{paragraph}{5}{2\parindent}{0ex plus 0.1ex minus 0.1ex}%
{0ex}{\normalfont\small\itshape}}%
}{}

\newcommand\role[1]{\unskip}
\newcommand\aucollab[1]{\unskip}
  
\@ifundefined{tsGraphicsScaleX}{\gdef\tsGraphicsScaleX{1}}{}
\@ifundefined{tsGraphicsScaleY}{\gdef\tsGraphicsScaleY{.9}}{}
\def\checkGraphicsWidth{\ifdim\Gin@nat@width>\linewidth
	\tsGraphicsScaleX\linewidth\else\Gin@nat@width\fi}

\def\checkGraphicsHeight{\ifdim\Gin@nat@height>.9\textheight
	\tsGraphicsScaleY\textheight\else\Gin@nat@height\fi}

\def\fixFloatSize#1{}%\@ifundefined{processdelayedfloats}{\setbox0=\hbox{\includegraphics{#1}}\ifnum\wd0<\columnwidth\relax\renewenvironment{figure*}{\begin{figure}}{\end{figure}}\fi}{}}
\let\ts@includegraphics\includegraphics

\def\inlinegraphic[#1]#2{{\edef\@tempa{#1}\edef\baseline@shift{\ifx\@tempa\@empty0\else#1\fi}\edef\tempZ{\the\numexpr(\numexpr(\baseline@shift*\f@size/100))}\protect\raisebox{\tempZ pt}{\ts@includegraphics{#2}}}}

\AtBeginDocument{\def\includegraphics{\@ifnextchar[{\ts@includegraphics}{\ts@includegraphics[width=\checkGraphicsWidth,height=\checkGraphicsHeight,keepaspectratio]}}}

\DeclareMathAlphabet{\mathpzc}{OT1}{pzc}{m}{it}

\def\URL#1#2{\@ifundefined{href}{#2}{\href{#1}{#2}}}

\def\UrlOrds{\do\*\do\-\do\~\do\'\do\"\do\-}%
\g@addto@macro{\UrlBreaks}{\UrlOrds}

\journal{}

\edef\fntEncoding{\f@encoding}

\makeatother

\newif\ifmultipleabstract\multipleabstractfalse%
\begin{document}

\begin{frontmatter}

\title{Efficient Computing Algorithm for High Dimensional Sparse Support Vector Machine
    \mbox{}}
%\tnotetext[mytitlenote]{The authors contribute equally to this paper}. 
%\cortext[myfootnote]{The authors contribute equally to this paper}
\author[mymainaddress]{Jiawei Wen}

% \cortext[mycorrespondingauthor]{Corresponding author}

\address[mymainaddress]{The Pennsylvania State University}
    
%\author{Jiawei Wen$^{a,*}$, Songshan Yang$^{a,*}$, Changcheng Li$^a$, Ethan X. Fang$^a$ and Runze Li$^a$ \fnref{myfootnote}}

%\address{$^a$The Pennsylvania State University at University Park\\}
% \fntext[myfootnote]{This research was supported by National Science Foundation Grants DMS 1820702, DMS 1953196 and DMS 2015539.}
%\cortext[myfootnote]{The authors contribute equally to this paper}
%\cortext[myfootnote]{The authors contribute equally to this paper}




\begin{abstract}
In recent years, considerable attention has been devoted to the regularization models due to the presence of high-dimensional data in scientific research. Sparse support vector machine (SVM) are useful tools in high-dimensional data analysis, and they have been widely used in the area of econometrics. Nevertheless, the non-smoothness of objective functions and constraints present computational challenges for many existing solvers in the presence of ultra-high dimensional covariates. In this paper, we design efficient and parallelizable algorithms for solving sparse SVM problems with high dimensional data through feature space split. The proposed algorithm is based on the alternating direction method of multiplier (ADMM). We establish the rate of convergence of the proposed ADMM method and compare it with existing solvers in various high and ultra-high dimensional settings. The compatibility of the proposed algorithm with parallel computing can further alleviate the storage and scalability limitations of a single machine in large-scale data processing. 
\end{abstract}

\end{frontmatter}
    
\section{Introduction}
\label{sec:intro}

Nowadays, enormous large-scale datasets have aroused frequently in many fields of modern scientific research, from genomics and biomedical science to finance and econometrics. This presents challenges to classic statistical methods that become ineffective or infeasible when the dimension $p$ is very large. More discussions on big data challenges in statistical analysis can be found in \citeay{jordan2013statistics} and \citeay{fan2014challenges}. In response to the advent of big data, studies on distributed learning for statistical estimation problems have drawn considerable attention in recent years \cite{zhang2012communication,wang2014median,battey2015distributed,lee2015communication,lin2017distributed,wang2017efficient,lin2018optimal,jordan2018communication}, where the focus is on the large $n$ problem and the data is partitioned across samples to a manageable size. In many scientific applications, collected datasets are common to have a huge number of features rather than samples, in which case it is more reasonable to split across the feature space. For example, \citeay{song2015split} devises a split-and-merge approach (SAM) for variable selection in high dimensional data analysis. SAM first partitions the ultrahigh dimensional dataset into a number of lower dimensional subsets, performs variable selection on each subset, and then performs a second Bayesian variable selection. One drawback is that when there exist strong correlations among predictors, the first step tends to select many correlated but irrelevant variables. Dealing with strong correlations among features is an inherent challenge of feature space partition. To address this issue for penalized regression models, \citeay{wang2016decorrelated}  proposes a parallel computing framework that performs one-step decorrelation before partition. 

In recent years, the alternating direction method of multipliers (ADMM) has drawn considerable attention due to its applicability to massive optimization problems. ADMM is proposed originally in \citeay{glowinski1975approximation} and \citeay{gabay1976dual}. \citeay{boyd2011distributed} gives a systematic review on this topic. In this paper, we adopt a specially designed 3-block ADMM algorithm as the building block in solving a class of optimization problems represented by sparse support vector machine,
\begin{equation}
\setlength{\abovedisplayskip}{3pt}
\setlength{\belowdisplayskip}{3pt}
\label{general_form}
\begin{aligned}
\hat{\bbeta} = \argmin_{\bbeta} L(\bbeta) + \lambda \sum_{j=1}^p  \alpha_j \vert \beta_j \vert
 \coloneqq \argmin_{\bbeta} L(\bbeta) + \lambda  \Vert \balpha \circ \bbeta\Vert_1,
\end{aligned}
\end{equation}
where $L(\bbeta)$ is a convex but non-smooth loss function,
$\bbeta=(\beta_1, \cdots, \beta_p)$ is an unknown feature coefficient vector, $\lambda$ is the penalty parameter, and $\balpha = (\alpha_1, \cdots, \alpha_p)^T$ is the weight parameter in the weighted $\ell_1$ penalty. When $\balpha = \bm{1}_p$, the weighted $\ell_1$ penalized problem reduces to the $\ell_1$ penalized problem. $\ell_1$ penalty tends to over-penalize large coefficients, and weighted $\ell_1$ penalty can address this issue by assigning different weights to coefficients. The choice of weight vector $\balpha$ plays a critical role in the performance of weighted $\ell_1$ penalty. In this work, we apply a two-step procedure to choose a proper $\balpha$. The first step computes an
$\ell_1$-penalized estimator as the initial estimator and the second step computes the final estimator using an adaptively chosen $\balpha$ dependent on the initial estimator.

Support vector machine \citep{cortes1995support} is an important tool used in binary classification, which has been used to model the data in the area of econometircs \cite{sunjiesvm,chaudhuri2014support,virag2013application,abedin2019topological}.  Consider a random sample of size $n$, $(\bx_i, y_i), i=1,\cdots,n$, with $\bx_i \in \mR^{p}$ and class label $y_i \in \{-1, 1\}$. The learning parameter is a $(p+1)$-vector $\bbeta^T=(\beta_0, (\bbeta^+)^T)$, where $\beta_0$ is the intercept term, and $\bbeta^+$ is the parameter vector without the intercept. The primary goal of linear SVM is to estimate a separating hyperplane $\bx^T\bbeta=0$ for two classes. 
Denote the feature matrix $\bX = (\bx_1, \bx_2, \cdots \bx_n)^T = (\bX_1, \cdots \bX_p)$ and labels $\by = (y_1, y_2, \cdots y_n)^T$. %There has been a rich literature in the study of support vector machines. 
The standard $\ell_2$-norm SVM estimates $\bbeta$ via solving the following optimization problem,
\begin{equation}
\setlength{\abovedisplayskip}{3pt}
\setlength{\belowdisplayskip}{3pt}
\min_{\bbeta} \frac{1}{n}\sum_{i=1}^n(1-y_i\bx_i^T\bbeta^+-y_i\beta_0)_{+} + \lambda_n \Vert \bbeta^+ \Vert_2^2,
\end{equation}
where the first term is the non-smooth empirical ``hinge'' loss. In high dimensional problems, the performance of the standard $\ell_2$-norm SVM can be adversely affected by irrelevant variables in the model. This motivates research on sparse SVM such as the $\ell_1$-norm SVM \citep{zhu20041},
\begin{equation}
\setlength{\abovedisplayskip}{3pt}
\setlength{\belowdisplayskip}{3pt}
\label{model_svm}
\min_{\bbeta} \frac{1}{n}\sum_{i=1}^n(1-y_i\bx_i^T\bbeta^+-y_i\beta_0)_{+} + \lambda_n \Vert \bbeta^+ \Vert_1.
\end{equation}
The $\ell_1$-norm SVM can be formulated as a linear program and solved by standard linear programming solver such as \emph{lpSolve} when data dimension is moderate.
%For nonconvex penalized SVM, we will compare with an R package called \emph{penalizedSVM} which supports more general penalties such as SCAD and elastic net. 

%While the aforementioned solvers function well in moderate dimensions, they may break down when dimension is higher than tens of thousands. 

In this paper we propose efficient and parallelizable algorithms for sparse SVM.
The proposed algorithms significantly improve the computing capacity by feature space splitting. The basic idea is to split feature space into smaller subsets, carry out computation independently on subproblems, and coordinate local solutions to update the global solution. In Section \ref{sec:SVM}, we present the distributed computational framework based on the 3-block ADMM for sparse SVM and derive the linear rate of convergence of the algorithm. %and re-visit the strong oracle property of the nonconvex penalized SVM estimator obtained by LLA.
In Section \ref{sec: exp}, we demonstrate the numerical and statistical efficiency of the proposed framework in high and ultra-high dimensional settings through a number of Monte Carlo simulations and real data analysis of a Chinese supermarket. 

Throughout the paper, we use the following notations. For $\bm{M} = (m_{ij})_{s\times t}$, $\Vert \bm{M} \Vert_{\max} = \max_{(i,j)} \vert m_{ij} \vert$, 
%is the entry-wise maximum absolute value, 
$\Vert \bm{M} \Vert_{\min} = \min_{(i,j) }\vert m_{ij} \vert$,
%is the entry-wise minimum absolute value.
$\lambda_{\min}(\bm{M})$ and $\lambda_{\max}(\bm{M})$ are the smallest and largest eigenvalues of $\bm{M}$, respectively. $\bX_{\bm{A}}$ is the sub-matrix of $\bX$ with the columns indexed by $\bm{A}$. For a positive semidefinite operator $\bm{P}$, $\Vert \bx \Vert_{\bm{P}}^2 = \bx^T \bm{P} \bx$.

\section{Review of ADMM}

ADMM was proposed originally in \cite{glowinski1975approximation, gabay1976dual} and later \cite{boyd2011distributed} gives a systematic review on this topic. 
The non-smoothness of the objective function in \eqref{model_svm} hinders an efficient application of gradient-based methods. ADMM is one of the decomposition-coordination procedures that naturally decouple the non-smooth parts in the computation. In this respect, we decentralize problem \eqref{model_svm} into the following constrained optimization problem,
\begin{equation}
\begin{aligned}
\label{2block:2}
&\min_{\bbeta,\bz}  L(\bz) + \lambda  \Vert \balpha \circ \bbeta \Vert_1\\
& \text{s.t.} \ \ \bz+\bX\bbeta=\by.
\end{aligned}
\end{equation}
The decentralization of variables transforms the problem into a natural candidate of two-block ADMM algorithm. One key ingredient in ADMM is the augmented Lagrangian function,
$$\mL_{\phi}(\bz, \bbeta; \bgamma) = L(\bz)+ \lambda \Vert \balpha \circ \bbeta\Vert_1 + \langle \bgamma, \bz+\bX\bbeta-\by \rangle+\frac{\phi}{2}\Vert \bz+\bX\bbeta-\by \Vert_2^2,$$ where $\bgamma \in \mR^n$ is the Lagrangian multiplier and $\phi>0$ is the parameter associated with the quadratic term. The classic iterative scheme at the iteration $k$ for two-block ADMM is given by
\begin{equation*}
\begin{aligned}
\bbeta^{k+1} &=\argmin_{\bbeta} \mL_{\phi}(\bbeta, \bz^k;\bgamma^k) \\
\bz^{k+1} &=\argmin_{\bz} \mL_{\phi}(\bbeta^{k+1}, \bz;\bgamma^k)\\
\bgamma^{k+1} &=\bgamma^{k}+\theta\phi(\bbeta^{k+1}+\bz^{k+1}-\by),\\
\end{aligned}
\end{equation*}
In this scheme, the primal variables $\bbeta$ and $\bz$ are updated alternatively via a single Gauss-Seidel pass. The dual variable $\bgamma$ is updated using the gradient ascent method, where $\theta$ is a tuning parameter controlling the step size. In convergence analysis, $\theta$ is often restricted to $(0, \frac{1+\sqrt{5}}{2})$. We favor a larger $\theta$ for faster convergence in practice, but not too large to retain theoretical convergence; thus a natural choice is to set $\theta = \frac{1+\sqrt{5}}{2}$. The convergence of the classic 2-block ADMM has been widely studied in literature. We refer to \cite{gabay1976dual, glowinski1975approximation, fortin2000augmented, goldfarb2012fast, douglas1956numerical, eckstein1990alternating, deng2016global} for some detailed discussions. One potential limitation is that the two-block formulation does not naturally parallelize variable updates and may suffer from memory-intensive operations such as large matrices multiplication. Major computational cost of two-block ADMM for solving \eqref{2block:2} comes from the $\bbeta$ update, which takes up to $O(np)$ operations and could impede an efficient execution of the algorithm in case $p$ is very large. To overcome this issue, we propose a new 3-block semi-proximal ADMM framework that capacitates a parallel update of $\bbeta$. 

Suppose we can split the feature matrix and coefficients into $K > 1$ parts,
\begin{equation*}
\begin{aligned}
&\bX = (\bX_1, \cdots, \bX_K),  \ \ \bbeta = \begin{pmatrix}
\bbeta_1\\
\bbeta_2\\
\vdots\\
\bbeta_K
\end{pmatrix},\ \ \bX\bbeta=\sum_{i=1}^K \bX_i\bbeta_i.
\end{aligned}
\end{equation*}
Then problem \eqref{2block:2} can be rewritten as a 3-block constrained optimization problem,
\begin{equation}
\begin{aligned}
\label{Quan:3block}
&\min_{\bbeta,\bz,\bomega} && L(\bz) + \sum_{i=1}^{K}\lambda \Vert \balpha_i \circ \bbeta_i \Vert_1,\\
&\text{s.t.} && \bX_1\bbeta_1+\bz+\bomega_2+\cdots+\bomega_K=\by,\\ 
& && \bomega_i=\bX_i\bbeta_i,\ \ i=2,\cdots,K.
\end{aligned}
\end{equation}
Intuitively, slack variables $\bomega_i, i=2,\cdots,K$ store information of each local update $\bbeta_i$. Note that although $\bbeta_i$ will be updated separately, they account for a single variable block in the problem. Likewise, all $\bomega_i$ together make up the third variable block. 
\begin{remark}
There may exist multiple ways to transform a problem into a form that ADMM can handle, and different formulations of the slack variables and constraints may give rise to different algorithms. For formulation \eqref{Quan:3block}, the role of $\bX_1\bbeta_1$ is not special and $\bX_i\bbeta_i, i=1,\cdots, K$ are exchangable. 
\end{remark}
The augmented Lagrangian function for \eqref{Quan:3block} is given by
\begin{equation}
\label{equ:lag}
\begin{aligned}
\mL{\phi}(\bbeta,\bz,\bomega;\bgamma)
%&=f(\bbeta_1,\cdots,\bbeta_K)+g(\bz)-<b,\bomega>+<\bgamma, F^T\by+G^T\bz+H^T\bomega-c>\\
%&+\frac{\phi}{2}\Vert F^T\by+G^T\bz+H^T\bomega-c \Vert^2\\
&=\frac{1}{n}\sum_{i=1}^n(1-y_i\bx_i^T\bbeta^+-y_i\beta_0)_{+}  + \lambda  \sum_{i=1}^{K}\Vert \balpha_i \circ \bbeta_i \Vert_1 \\
& +\bgamma_1^T(\bX_1\bbeta_1+\bz+\bomega_2+\cdots+\bomega_K-\by) +\sum_{i=2}^K\bgamma_i^T(\bX_i\bbeta_i - \bomega_i) \\
& + \frac{\phi}{2} \Vert \bX_1\bbeta_1+\bz+\bomega_2+\cdots+\bomega_K - \by \Vert_2^2 + \frac{\phi}{2} \sum_{i=2}^K\Vert \bX_i\bbeta_i-\bomega_i\Vert_2^2.
\end{aligned}
\end{equation}
As we can see from \eqref{equ:lag}, each $\bbeta_i$ is decoupled in the quadratic term, which allows a natural parallelization for $\bbeta$ updates.

Two-block ADMM can be directly extended for solving \eqref{Quan:3block}. At iteration $k$, it alternatingly minimizes the augmented Lagrangian between $\bbeta$ and $\bz$ and is referred to as Gauss-Seidel multi-block ADMM,
\begin{equation}
\label{alg: ADMM3b}
\begin{cases}
\bbeta^{k+1} &= \argmin \mL_{\phi}(\bbeta, \bz^k, \bomega^k; \bgamma^k)\\
\bz^{k+1} &=\argmin \mL_{\phi}(\bbeta^{k+1}, \bz, \bomega^k; \bgamma^k)\\
\bomega^{k+1} &=\argmin \mL_{\phi}(\bbeta^{k+1}, \bz^{k+1}, \bomega; \bgamma^k) \\
%= (H^TH)^{-1}H(\by-F \bbeta^{k+1}-G \bz^{k+1}),\\
\bgamma_1^{k+1}&=\bgamma_1^k+\theta\phi(\bX_1\bbeta_1^{k+1}+\bz^{k+1}+\sum_{i=2}^K\bomega_i^{k+1}-\by)\\
\bgamma_i^{k+1}&=\bgamma_i^k+\theta\phi(\bX_i\bbeta_i^{k+1}-\bomega_i^{k+1}), \ \ i = 2,\cdots,K.
\end{cases}
\end{equation}
Procedure \eqref{alg: ADMM3b} functions well in many practical cases; however, its theoretical convergence has remained unclear for long. In fact, it was shown by \cite{chen2016direct} that Gauss-Seidel multi-block ADMM is not necessarily convergent. Recently, \cite{sun2015convergent} proposes a convergent symmetric Gauss-Seidel based semi-proximal ADMM for convex programming problems with three separable blocks and the third part being linear. It takes a special block coordinate descent cycle for updating each block and update the third block variable twice. We adopt their procedure and obtain the following update scheme for \eqref{Quan:3block},
\begin{equation}
\label{alg: sPADMM3c}
\begin{cases}
\bbeta^{k+1} &= \argmin \mL_{\phi}(\bbeta, \bz^k, \bomega^k; \bgamma^k) + \frac{\phi}{2}\Vert \bbeta-\bbeta^k\Vert^2_{\mT_{f}}\\
\bomega^{k+\frac{1}{2}} &= \argmin \mL_{\phi}(\bbeta^{k+1}, \bz^k, \bomega; \bgamma^k)\\ 
%= (H^TH)^{-1}H(\by-F \bbeta^{k+1}-G \bz^k),\\
\bz^{k+1} &=\argmin \mL_{\phi}(\bbeta^{k+1}, \bz, \bomega^{k+\frac{1}{2}}; \bgamma^k) + \frac{\phi}{2}\Vert \bz-\bz^k \Vert^2_{\mT_{g}}\\
\bomega^{k+1} &=\argmin \mL_{\phi}(\bbeta^{k+1}, \bz^{k+1}, \bomega; \bgamma^k) \\
%= (H^TH)^{-1}H(\by-F \bbeta^{k+1}-G \bz^{k+1}),\\
\bgamma_1^{k+1}&=\bgamma_1^k+\theta\phi(\bX_1\bbeta_1^{k+1}+\bz^{k+1}+\sum_{i=2}^K\bomega_i^{k+1}-\by)\\
\bgamma_i^{k+1}&=\bgamma_i^k+\theta\phi(\bX_i\bbeta_i^{k+1}-\bomega_i^{k+1}), \ \ i = 2,\cdots,K,
\end{cases}
\end{equation}
A major difference of \eqref{alg: sPADMM3c} from the classic semi-proximal ADMM is that \eqref{alg: sPADMM3c} performs an extra intermediate step to compute $\bomega^{k+\frac{1}{2}}$ before computing $\bz^{k+1}$. Therefore, the practical success and efficiency of the algorithm relies heavily on the computational cost to update $\bomega$ and we will show that this extra cost is negligible in our case. $\mT_{f}$ and $\mT_{g}$ are two optional self-adjoint positive semidefinite operators. One desirable reason for including $\mT_{f}$ and $\mT_{g}$ is to ensure that $\{\bbeta^{k+1}\}$ and $\{\bz^{k+1}\}$ are well defined. A general principle is that $\mT_{f}$ and $\mT_{g}$ should be as small as possible, while the optimization problems are still easy to compute. %In this study, we simply take $\mT_f=\mT_g=0$ so that the proximal terms $\Vert x-x^k\Vert^2_{\mT_f}$ and $\Vert y-y^k \Vert^2_{\mT_g}$ are absent. 
The effect of tuning parameter $\theta$ on the algorithm convergence has been discussed in a number of works including \cite{fazel2013hankel} and \cite{fortin2000augmented}, where algorithms convergences are established when $\theta$ is constrained to $(0, \frac{1+\sqrt{5}}{2})$. In our numerical experiments, we set $\theta = \frac{1+\sqrt{5}}{2}$. 

\section{Sparse Linear Support Vector Machines}
\label{sec:SVM}
In this section, we apply the proposed algorithm to sparse linear support vector machine. We discuss 3-block ADMM approach for the weighted $\ell_1$-SVM problem with a given weight vector $\balpha$, and then propose the two-step procedure for adaptively weighted $\ell_1$-SVM.
%Consider the binary classification problem, and let $(Y_i, \bx_i)_{i=1}^n$ be a random sample. Feature matrix $\bX = (X_{0}, \bX_+) = ( \bx_{1}, \cdots, \bx_{n})^T$, where $X_{0}$ represents the intercept term. 
%The separating hyperplane $\bX\bbeta^*$ is parameterized by $\bbeta^* = (\beta_0^*, (\bbeta^{+*})^T)^T,$ which is the minimizer of the population hinge loss, i.e.,
%\begin{equation}
%    \bbeta^* = \argmin_{\bbeta}  E(1-Y\bX \bbeta)_+.
%\end{equation}

\subsection{ADMM Algorithms for Weighted $\ell_1$ Penalized SVM}

The weighted $\ell_1$-penalized SVM estimation problem is given by
\begin{equation}
\setlength{\abovedisplayskip}{3pt}
\setlength{\belowdisplayskip}{3pt}
\label{svm_model}
\min_{\bbeta} \frac{1}{n}\sum_{i=1}^n(1-y_i\bx_i^T\bbeta^+-y_i\beta_0)_{+} + \lambda \Vert \balpha \circ \bbeta^+ \Vert_1.
\end{equation}

%\subsection{Weighted \texorpdfstring{$\ell_1$}--penalized SVM}
We first outline the algorithm for the weighted $\ell_1$-penalized SVM. Divide the learning parameter $\bbeta^+$ and the matrix $\by \bX^T$ into $G$ blocks
\begin{equation*}
\setlength{\abovedisplayskip}{3pt}
\setlength{\belowdisplayskip}{3pt}
\bbeta^T=(\beta_0, (\bbeta^+)^T)=(\beta_0, \bbeta_1^T, \cdots, \bbeta_G^T)
\end{equation*}
\begin{equation*}
\bA=\begin{pmatrix}
y_1\bx_1^T\\
y_2\bx_2^T\\
\vdots\\
y_n\bx_n^T\\
\end{pmatrix}=\begin{pmatrix}
\bA_0, \bA_1, \bA_2, \cdots, \bA_G
\end{pmatrix},
\end{equation*}
where $\bbeta_g \in \mR^{p_g}$, $g=1,\cdots,G$, $\bm{A}_0 = \by$ and $\bm{A}_g \in \mR^{n\times p_g}, \sum_{g=1}^G p_g = p$. Then problem \eqref{svm_model} can be reformulated as a 3-block constraint minimization problem:
\begin{equation}
\setlength{\abovedisplayskip}{3pt}
\setlength{\belowdisplayskip}{3pt}
\label{svm:3block}
\begin{aligned}
&  \min_{\bbeta,\bz,\bomega} && \sum_{g=1}^G \lambda \Vert \balpha_g \circ \bbeta_g \Vert_1 + \frac{1}{n} \bm{1}_n^T(\bz)_+\\
&\text{s.t.} && \bz+\sum_{g=1}^G \bomega_g+\bA_0 \beta_0=\bm{1}_n, \bA_g\bbeta_g=\bomega_g, \ \ g=1,\cdots G.
\end{aligned}
\end{equation}
%\begin{empheq}[left=\empheqlbrace]{align}
Through a similar discussion of the quantile regression and LPD in \cite{WEN2023105426} and \cite{WenLPD}, $\bbeta$ updates consist of the following problems:
\begin{equation}
\setlength{\abovedisplayskip}{3pt}
\setlength{\belowdisplayskip}{3pt}
 \begin{aligned}
\label{betai.4}
\beta_0^{k+1}&=\frac{1}{\Vert \by \Vert^2} \by^T(\bm{1}_n-\bz^k-\sum_{g=1}^G\bomega_g^k-\frac{\bgamma_0^k}{\phi}),\\
\bbeta_g^{k+1}&=\argmin_{\bbeta_g \in \mR^{p_g}} \lambda \Vert \balpha_g \circ \bbeta_g \Vert_1 + \frac{\phi}{2}\Vert \bm{A}_g\bbeta_g-\bomega_g^k+\frac{\bgamma_g^k}{\phi} \Vert^2_2.
\end{aligned}   
\end{equation}
By introducing proximal terms $\bm{\mT_g} = \eta_g\bm{I}_{p_g}-\phi\bm{A}_g^T\bA_i$ with $\eta_g > \phi\lambda_{\text{max}}(\bm{A}_g^T\bm{A}_g)$ to subproblem \eqref{betai.4}, we obtain the proximal update given by
%\label{beta_prox}
%\bbeta_i^{k+1}=\argmin_{\bbeta_i \in \mR^{p_i}} \lambda \Vert \balpha_i \circ \bbeta_i \Vert_1 + \frac{\phi}{2}\Vert \bm{A}_i\bbeta_i-\bomega_i^k+\frac{\bgamma_i^k}{\phi} \Vert^2 +  \frac{\phi}{2}\Vert \bbeta_1 - \bbeta_1^k  \Vert^2_{\bm{\mT_i}}, \ \ i = 2,\cdots,G,\\
%\end{equation} $\bm{\mT_i} = \eta_i\bm{I}_{p_i}-\bm{A}_i^T\bA_i$ with $\eta_i > \lambda_{\text{max}}(\bm{A}_i^T\bm{A}_i)$ and $\lambda_{\text{max}}(\cdot)$ is the largest eigenvalue of matrix. Then \eqref{beta_prox} endorses a closed form solution given by 

\begin{equation}
\begin{aligned}
\label{svm_beta_i}
\bbeta_g^{k+1}%&=\argmin_{\bbeta_i \in \mR^{p_i}} \lambda \Vert \balpha_i \circ \bbeta_i\Vert_1 \\
%& \quad + \frac{\eta_i}{2}\Vert \bbeta_i - \frac{\eta_i\bbeta_i^k - \phi \bm{A}_i^T(\bm{A}_i\bbeta_i-\bomega_i^k+\frac{\bgamma_i^k}{\phi})}{\eta_i})\Vert_2^2\\
&= \text{S}\Big(\bbeta_{g}^k - \frac{\phi}{\eta_g}\bA_{g}^T(\bm{A}_g\bbeta_g-\bomega_g^k+\frac{\bgamma_g^k}{\phi}), \frac{\alpha_g \lambda}{\eta_g} \Big)_{j=1,\cdots,p_g}.\\
\end{aligned} 
\end{equation}
The update for $\bz$ endorses a closed form solution as follows,
\begin{equation}
\setlength{\abovedisplayskip}{3pt}
\setlength{\belowdisplayskip}{3pt}
\label{zupdate.4}
\begin{aligned}
\bz^{k+1}%&=\argmin_{\bz} \frac{1}{n} \bm{1}_n^T(\bz)_+ + \bgamma_0^T\bz +\frac{\phi}{2} \Vert \by\beta_0^{k+1}+\bz+\sum_{g=1}^G\bomega_i^{k+\frac{1}{2}}-\bm{1}_n \Vert^2\\
 = \Big(\bm{1}_n - \by\beta_0^{k+1} - \sum_{g=1}^G \bomega_g^{k+\frac{1}{2}}-\frac{\bgamma_0^k}{\phi} - \frac{1}{n\phi} \Big)_+ 
   - \Big(\bm{1}_n-\by\beta_0^{k+1}-\sum_{g=1}^G\bomega_g^{k+\frac{1}{2}} -\frac{\bgamma_0^k}{\phi} \Big)_- 
 \end{aligned}
\end{equation}
Given $\bbeta^{k+1}$ and $\bz^{k+1}$, other variables are updated as follows,
\begin{align}
\label{shalf.4}
\bomega_g^{k+\frac{1}{2}}=&\frac{1}{G+1}(\bm{1}_n-\by\beta_0^{k+1}-\bz^k+(G+1)\bm{A}_g\bbeta_g^{k+1}-\sum_{g=1}^G \bm{A}_g\bbeta_g^{k+1})\\
\label{supdate.4}
\bomega_g^{k+1} =& \frac{1}{G+1}(\bm{1}_n-\by\beta_0^{k+1}-\bz^{k+1}+(G+1)\bm{A}_g\bbeta_g^{k+1}-\sum_{g=1}^G \bm{A}_g\bbeta_g^{k+1})\\
\label{gamma0.4}
\bgamma_0^{k+1}=&\bgamma_0^k+\theta\phi(\by\beta_0^{k+1}+\bz^{k+1}+\sum_{g=1}^G \bomega_g^{k+1}-\bm{1}_n)\\
\label{gammai.4}
\bgamma_g^{k+1}=&\bgamma_g^k+\theta\phi(\bm{A}_g\bbeta_g^{k+1}-\bomega_g^{k+1}), \ \ g = 1,\cdots,G.
%\end{empheq}
\end{align}
We summarize algorithms for SVM as in Algorithm \ref{alg:svm-cd} and \ref{alg:svm_prox}. We present the convergence rate for ADMM-prox under the SVM setting in Theorem \ref{prop:svm-prox}. The proof is very similar to \cite{WEN2023105426} and we omit it for brevity.

\begin{algorithm}[hbtp]
\caption{ADMM-CD for weighted $\ell_1$-norm SVM}
\label{alg:svm-cd}
\begin{algorithmic}
\REQUIRE ~ $\bbeta^0, \bomega^0, \bz^0, \bgamma^0$, and $\phi >0, \theta >0$ are given.
\WHILE{the stopping criterion is not satisfied,}
\STATE Compute $\bbeta^{k+1}$ by \eqref{betai.4};
\STATE Compute $\bomega^{k+\frac{1}{2}}$ by \eqref{shalf.4};
\STATE  Compute $\bz^{k+1}$ by \eqref{zupdate.4};
\STATE  Compute $\bomega^{k+1}$ by \eqref{supdate.4};
\STATE Update $\bgamma^{k+1}$ by \eqref{gamma0.4} and \eqref{gammai.4}.
\ENDWHILE 
\end{algorithmic}
\end{algorithm}

\begin{algorithm}
\caption{ADMM-prox for weighted $\ell_1$-norm SVM}
\label{alg:svm_prox}
\begin{algorithmic}
\REQUIRE ~ $\bbeta^0, \bomega^0, \bz^0, \bgamma^0$, and $\phi >0, \theta >0$ are given.
\WHILE{the stopping criterion is not satisfied,}
\STATE Compute $\bbeta^{k+1}$ by \eqref{svm_beta_i};
\STATE Compute $\bomega^{k+\frac{1}{2}}$ by \eqref{shalf.4};
\STATE  Compute $\bz^{k+1}$ by \eqref{zupdate.4};
\STATE  Compute $\bomega^{k+1}$ by \eqref{supdate.4};
\STATE Update $\bgamma^{k+1}$ by \eqref{gamma0.4} and \eqref{gammai.4}.
\ENDWHILE 
\end{algorithmic}
\end{algorithm}

%We present the result of convergence rate in Theorem \ref{prop:svm-prox} for Algorithm \ref{alg:svm_prox}. The proof of Theorem \ref{prop:svm-prox} can be derived similarly as Theorem \ref{prop:quan_prox} and we omit it for brevity.
%We establish the convergence properties of Algorithm \ref{alg:qreg-prox}. .
\begin{theorem}
\label{prop:svm-prox}%Suppose that $(\bX, \by)$, where $\bX=(\bx_1,\cdots,\bx_n)^T, \by=(y_1, \cdots, y_n)^T,\\ i=1,\cdots,n$ are in general positions and the solution of \eqref{Quan:multi-block} is unique. Let $f, g, h, F, G, H$ be functions and operators defined in \eqref{Quan:operator}. Then under the condition that $\theta \in (0, (1+\sqrt{5})/2)$, the sequence ${(\bbeta^k, \bz^k, \bomega^k, \bgamma^k)}$ generated by Algorithm \ref{alg:qreg-prox} converges in probability to a unique limit $(\bar{\bbeta}, \bar{\bz}, \bar{\bomega}, \bar{\bgamma})$ with $(\bar{\bbeta}, \bar{\bz}, \bar{\bomega})$ solving \eqref{Quan:3block} and $\bar{\bgamma}$ is the dual optimal. 
For $\theta \in (0, (1+\sqrt{5})/2)$, the sequence ${(\bbeta^k, \bz^k, \bomega^k, \bgamma^k)}$ generated by Algorithm~\ref{alg:svm_prox} converges to a limit point ${(\bar{\bbeta}, \bar{\bz}, \bar{\bomega}, \bar{\bgamma}})$ that solves the optimization problem \eqref{svm:3block} and its dual problem. Furthermore, there exists a constant $\mu \in (0,1)$ such that 
$$\text{Dist}^{k+1} \leq \mu Dist^{k},$$ 
where $\text{Dist}^k$ is defined as
\begin{equation}
\begin{aligned}
\text{Dist}^{k} =& \sum_{g=0}^G \left( \Big \Vert \bA_g(\bbeta_g^{k}- \bar{\bbeta}_g) - \frac{1}{G+1}\bA  (\bbeta^{k}- \bar{\bbeta}) \Big \Vert_2^2 + \Vert \bbeta_g^{k} - \bar{\bbeta}_g \Vert^2_{\bm{\mT_g}} \right) \\
 &+ \sum_{g=0}^G \Vert \bbeta_g^{k} - \bar{\bbeta}_g \Vert^2_{\bm{\mT_g}} +  \Vert \bz^{k} -\bar{\bz}  \Vert^2_2 + \frac{G}{G+1} \Vert \bz^{k}- \bz^{k-1}\Vert^2_2 \\
 & + \frac{m_1}{G+1} \Vert \sum_{g=0}^G \bA_g(\bbeta_g^{k}-\bar{\bbeta}_g) + (\bz^{k} -\bar{\bz}) \Vert_2^2,
\end{aligned}
\end{equation}
with $m_1 = 1+d_1-d_1\theta-(1-d_1)\min\{\theta, \frac{1}{\theta}\}$ and $d_1 \in (0, \frac{1}{2})$. 
\end{theorem}

%\subsection{Sparse SVM with Nonconvex Penalty}
%In this part, we first revisit the strong oracle property of the folded concave penalized SVM estimator obtained by one-step LLA and then 

%To generalize ADMM-CD and ADMM-prox to the nonconvex penalized SVM with the help of LLA.
%\citeay{zhang2016variable} establishes the oracle property of the nonconvex penalized SVM and shows that with probability tending to one, the LLA algorithm is able to find the oracle estimator in one step with an appropriate initial estimator. %As a complementary work to \citeauthor{zhang2016variable} (\citeyear{zhang2016variable}), we provide non-asymptotic probability bounds of the one-step LLA reaching the oracle estimator when initialized by Lasso estimator and present it in the supplemental material.

\subsection{Two-step Procedures for Adaptively Weighted $\ell_1$ Penalized SVM}
\label{sec:svm_alg}

We adopt the two-step procedure to choose $\balpha$ adaptively for the weighted $\ell_1$ penalized SVM, and the corresponding algorithms are summarized in Algorithms~\ref{alg:scad_svm_cd} and \ref{alg:scad_svm_prox} for two-step ADMM-CD and two-step ADMM-prox, respectively. Note that the two-step procedure is essentially the ADMM computational approach combined with the one-step LLA algorithm, and the oracle property of the one-step LLA algorithm under the sparse penalized SVM setting has been established by \citeay{zhang2016variable}.
Together with Theorem~\ref{prop:svm-prox}, the two-step procedure here should also enjoy oracle properties with a good convergence rate.
%As a complementary work to \citeay{zhang2016variable}, we provide non-asymptotic probability bounds of the one-step LLA reaching the oracle estimator when initialized by Lasso estimator and present it in the supplemental material.

\section{Numerical Experiments}
\label{sec: exp}
In this section, we evaluate the performance of the proposed algorithms on synthetic datasets and real data. For all ADMM-based methods, we implement the warm-start technique introduced in \citeay{friedman2007pathwise} and \citeay{friedman2010regularization}, which uses the solution from the previous $\lambda$ to initialize computation at the current $\lambda$. The stopping criterion of ADMM-based algorithms is provided in the 
Appendix.

\begin{algorithm}[H]
\caption{Two-step ADMM-CD for Sparse Penalized SVM}
\label{alg:scad_svm_cd}
\begin{algorithmic}
\REQUIRE ~ $\tilde{\bbeta}^{0}, \lambda, \upsilon, \tilde{\bz}^{0}, \tilde{\bgamma}^{0},  \tilde{\bomega}_i^{0}$, and $\phi >0, \theta =1.618, k=0$.
\WHILE{the stopping criterion is not satisfied,}
\STATE Update $\tilde{\bbeta}^{k+1}$ by solving 
\begin{equation*}
\begin{aligned}
\tilde{\bbeta}_0^{k+1} &=\frac{1}{\Vert \by \Vert^2} \by^T(\bm{1}_n-\bz^k-\sum_{g=1}^G \bomega_g^k-\frac{\bgamma_0^k}{\phi}),\\
\tilde{\bbeta}_g^{k+1} &=\argmin_{\bbeta_g \in \mR^{p_g}} \upsilon \lambda \Vert \bbeta_g \Vert_1 + \frac{\phi}{2}\Vert \bm{A}_g\bbeta_g-\bomega_g^k+\frac{\bgamma_g^k}{\phi} \Vert^2_2, \ \ g = 1,\cdots,G\\
\end{aligned}
\end{equation*}
via coordinate descent algorithm;
\STATE Compute $\tilde{\bomega}^{k+\frac{1}{2}}$ by \eqref{shalf.4};
\STATE  Compute $\tilde{\bz}^{k+1}$ by \eqref{zupdate.4};
\STATE  Compute $\tilde{\bomega}^{k+1}$ by \eqref{supdate.4};
\STATE Update $\tilde{\bgamma}^{k+1}$ by \eqref{gamma0.4} and \eqref{gammai.4}.
\ENDWHILE \ The solution is denoted as $\hat{\bbeta}^{\ell_1}, \hat{\bz}^{\ell_1}, \hat{\bomega}^{\ell_1}$
\REQUIRE $\hat{\bbeta}^{0} = \hat{\bbeta}^{\ell_1}, \hat{\bz}^{0} = \hat{\bz}^{\ell_1},  \hat{\bomega}^{0}=\hat{\bomega}^{\ell_1}$ and $\phi >0, \theta =1.618, k=0$.
\WHILE{the stopping criterion is not satisfied,}
\STATE Compute $\alpha_j^{k+1} = \lambda^{-1}P^{'}_{\lambda}( \vert \hat{\beta_j}^{(k)} \vert)$ for $j=1,\cdots,p$;
\STATE Update $\hat{\bbeta}^{k+1}$ by \eqref{betai.4};
\STATE Compute $\hat{\bomega}^{k+\frac{1}{2}}$ by \eqref{shalf.4};
\STATE  Compute $\hat{\bz}^{k+1}$ by \eqref{zupdate.4};
\STATE  Compute $\hat{\bomega}^{k+1}$ by \eqref{supdate.4};
\STATE Update $\hat{\bgamma}^{k+1}$ by \eqref{gamma0.4} and \eqref{gammai.4}.
\ENDWHILE 
\end{algorithmic}
\end{algorithm}

\begin{algorithm}[H]
\caption{Two-step ADMM-prox for Sparse Penalized SVM}
\label{alg:scad_svm_prox}
\begin{algorithmic}
\REQUIRE ~ $\tilde{\bbeta}^{0}, \lambda, \upsilon, \tilde{\bz}^{0}, \tilde{\bgamma}^{0},  \tilde{\bomega}_i^{0}$, and $\phi >0, \theta =1.618, k=0$.
\WHILE{the stopping criterion is not satisfied,}
\STATE Update $\tilde{\bbeta}^{k+1}$ by 
\begin{equation*}
\begin{aligned}
\tilde{\bbeta}_0^{k+1}&=\frac{1}{\Vert \by \Vert^2} \by^T(\bm{1}_n-\bz^k-\sum_{g=1}^G \bomega_g^k-\frac{\bgamma_0^k}{\phi})\\
\tilde{\bbeta}_g^{k+1} &=\text{Shrink}\Big(\bbeta_{gj}^k - \frac{1}{\eta_g}\bA_{g(j)}^T(\bm{A}_g\bbeta_g-\bomega_g^k+\frac{\bgamma_g^k}{\phi}), \frac{\upsilon \lambda}{\phi\eta_g} \Big)_{j=1,\cdots,p_g};\\
\end{aligned}
\end{equation*}
\STATE Compute $\tilde{\bomega}^{k+\frac{1}{2}}$ by \eqref{shalf.4};
\STATE  Compute $\tilde{\bz}^{k+1}$ by \eqref{zupdate.4};
\STATE  Compute $\tilde{\bomega}^{k+1}$ by \eqref{supdate.4};
\STATE Update $\tilde{\bgamma}^{k+1}$ by \eqref{gamma0.4} and \eqref{gammai.4}.
\ENDWHILE \ The solution is denoted as $\hat{\bbeta}^{\ell_1}, \hat{\bz}^{\ell_1}, \hat{\bomega}^{\ell_1}$
\REQUIRE $\hat{\bbeta}^{0} = \hat{\bbeta}^{\ell_1}, \hat{\bz}^{0} = \hat{\bz}^{\ell_1},  \hat{\bomega}^{0}=\hat{\bomega}^{\ell_1}$ and $\phi >0, \theta =1.618, k=0$.
\WHILE{the stopping criterion is not satisfied,}
\STATE Compute $\alpha_j^{k+1} = \lambda^{-1}P^{'}_{\lambda}( \vert \hat{\beta_j}^{(k)} \vert)$ for $j=1,\cdots,p$;
\STATE Compute $\tilde{\bomega}^{k+\frac{1}{2}}$ by \eqref{shalf.4};
\STATE  Compute $\tilde{\bz}^{k+1}$ by \eqref{zupdate.4};
\STATE  Compute $\tilde{\bomega}^{k+1}$ by \eqref{supdate.4};
\STATE Update $\tilde{\bgamma}^{k+1}$ by \eqref{gamma0.4} and \eqref{gammai.4}.
\ENDWHILE 
\end{algorithmic}
\end{algorithm}

\subsection{Synthetic Study for SVM}
We benchmark the performances of ADMM-CD and ADMM-prox against the standard LP solver \emph{lpSolve} on a synthetic dataset. We generate $\bX \sim \bm{MN}_p(\bm{0}, \bSigma)$, where $\bSigma = (\sigma_{ij})$ with %nonzero elements $\sigma_{ii}$ = 1 for $i = 1, 2,\cdots, p$ and
$\sigma_{ij}=0.4^{\vert i-j \vert}$ for $1 \leq i, j \leq p$, $P(Y = 1) = \Phi(\bX\bbeta)$ where $\Phi(\cdot)$ is the CDF of the standard normal distribution. Let $q = 4$ and true active set $\mA = \{ 50, 1000, 1500, 2000 \}$. Let $\bbeta_j^+ = 1.1$ for $j \in \mA$. We set $(n,p)= (300, 3000)$ and $(300, 50000)$. Regularization parameter $\lambda$ is chosen by the $\text{SVMIC}_H$ criterion proposed in \citeay{zhang2016consistent}, 
\begin{equation}
\setlength{\abovedisplayskip}{3pt}
\setlength{\belowdisplayskip}{3pt}
\begin{aligned}
    \text{SVMIC}_H(\lambda_n) = \sum_{i=1}^n (1-y_i\hat{\beta}_0(\lambda_n)-y_i\bx_{i}^T\hat{\bbeta}_{+}(\lambda_n))_{+} 
     + L_n\vert \hat{\mA}(\lambda_n)\vert \log(n),
\end{aligned}
\end{equation}
where $\hat{\mA}(\lambda_n) = \{ \hat{\bbeta}_j(\lambda_n)\neq 0, j=1,\cdots p\}$. As suggested by \citeay{zhang2016consistent}, we choose $L_n = \log(\log(n))$ and select the $\lambda$ that minimizes $\text{SVMIC}_H$. We also generate an independent dataset of the same sample size to evaluate the test performance. We evaluate the algorithms performances averaged over $500$ replications by the following criteria.
\begin{itemize}[nosep]
    \item Test error: testing misclassification error rate.
    \item Signal: the number of selected relevant features, i.e., $\hat{\beta}_j \neq 0$ with $j \in \{50, 1000, 1500, 2000\}$.
    \item Noise: the number of selected irrelevant features, i.e., $\hat{\beta}_j \neq 0$ with $j \notin \{50, 1000, 1500, 2000\}$.
    \item AAC: the absolute value of the sample correlation between $\bX\bbeta^*$ and $\bX\hat{\bbeta}$. 
  %  \item Time: overall time (in seconds) of model fitting over the same tuning sequence of 10 $\lambda$ values.
\end{itemize}
AAC is an accuracy measure used in \citeay{cook2007dimension}. An AAC close to $1$ implies that the estimated direction matches that of Bayes rule. % We conduct $500$ replications to measure the variations of results. 
We denote the solution of \emph{lpSolve} by $\hat{\bbeta}^{\text{LP}}$ and we expect the sequence generated by the ADMM-based algorithm to be close to the solution provided by linear programming. %In figure \ref{fig:svm}, we plot the $\ell_2$ difference $\Vert \hat{\bbeta} - \hat{\bbeta}^{\text{LP}} \Vert_2^2$ against iterations averaged over $100$ replications for $\ell_1$-ADMM-CD and $\ell_1$-ADMM-prox, respectively, and the results align with our expectation. 
We summarize the evaluation performances of three methods in Table \ref{sim:svm}. Two-step adaptive SVM improves $\ell_1$-norm SVM on test accuracy, signal selection and AAC. When $p=3000$, ADMM-CD, ADMM-prox and \emph{lpSolve} have very similar performances. When $p$ is increased to $50000$, ADMM-CD and ADMM-prox result in better AAC than \emph{lpSolve}. The computational efficiency of \emph{lpSolve} is severely affected by the high dimensions and two-step \emph{lpSolve} fails due to excessive memory usage.

% the memory is 4gb.
\begin{table}[htbp]

	\begin{center}
		\caption{\label{sim:svm} Numerical comparisons for sparse SVM.}
		\scalebox{0.95}{
			\begin{tabular}{ccccc}
\multicolumn{5}{c}{$n=300,  p=3000$} \\
			\hline
			
 & Test error &  Signal &  Noise &  AAC\\   %  & Time
					     \hline

$\ell_1$-ADMM-CD  & 0.16 (0.00) &  3.99 (0.00) & 1.38 (0.02) &  0.97 (0.00) \\ %\\ & 172.1 (2.0)

$\ell_1$-ADMM-prox  & 0.16 (0.00) & 3.98 (0.00) & 1.26 (0.02)  &  0.97 (0.00) \\% \\%& 79.5 (1.7)

$\ell_1$-\emph{lpSolve}  & 0.16 (0.00) & 3.99 (0.00) & 0.94 (0.04) & 0.97 (0.00) \\   %& 24.2 (0.1) 

Two-step ADMM-CD  & 0.15 (0.00) & 4.00 (0.00) & 1.23 (0.02) & 0.99 (0.00) \\ %  & 195.2 (1.3)

Two-step ADMM-prox & 0.16 (0.00) & 3.99 (0.00)  & 1.53 (0.03) & 0.97 (0.00) \\ %& 184.2 (1.4)

Two-step \emph{lpSolve} &  0.16 (0.00) & 4.00 (0.00) &0.95 (0.04) & 0.98 (0.00) \\ %\\% & 66.4 (0.1) 
		 \hline
\multicolumn{5}{c}{$n=300,  p=50000$} \\
			\hline
 & Test error &  Signal &  Noise &  AAC  \\  % & Time
					     \hline

$\ell_1$-ADMM-CD & 0.17 (0.00) &  3.95 (0.01) &  1.41 (0.02) & 0.96 (0.00) \\ %  \\  & 324.4 (3.1)

$\ell_1$-ADMM-prox & 0.17 (0.00) & 3.96 (0.01) & 1.28 (0.03)  & 0.96 (0.00) \\ % & 287.6 (9.4) 

$\ell_1$-\emph{lpSolve}  & 0.17 (0.00) & 3.95 (0.01)  &  0.90 (0.04) & 0.95 (0.00) \\  % & 949.4 (20.3) 
   
Two-step ADMM-CD  & 0.17 (0.00) & 3.98 (0.00) & 1.45 (0.03) & 0.96 (0.00)  \\ % & 1909.1 (20.5) 

Two-step ADMM-prox  & 0.16 (0.00) & 3.95 (0.01) & 1.70 (0.03) & 0.96 (0.00)  \\ % & 1781.3 (19.3)

Two-step \emph{lpSolve}  & &    \ding{55} & & \\ 
		 \hline

			\end{tabular}
			}
	\end{center}
\end{table}

%\subsection{Real data examples}

\subsection{Gene Expression Cancer RNA-Seq Dataset}
%\{change this real data}
The dataset comes from the UCI machine learning repository, and the original dataset is maintained by the cancer genome atlas pan-cancer analysis project. There are 801 samples and 20531 variables for each sample. Each variable is an RNA-Seq gene expression level measured by illumina HiSeq platform. Breast Carcinoma (BRCA) is one type of tumor, and we apply penalized support vector machine to identify samples that carry BRCA. As a prepossessing step, we delete genes that have constant expression levels across all samples, and use the remaining 20260 genes as candidate predictors. We randomly split the dataset into a $500$ training sample and $301$ test sample. The procedure is replicated for $100$ times, and the performances are reported in Table \ref{realdata:svm}. 

We compare different models and algorithms on the proportion of misclassified instances in the training and testing data, the model size and the computational time. The regularization parameter is selected by cross-validation. We can observe that under $\ell_1$ penalty, ADMM-CD has similar performance to the \emph{lpSolve} solution, with computation time 30 times less than that of the latter's. ADMM-prox performs slightly worse than ADMM-CD, with a larger active set identified. We also implement two-step penalized SVM and compare it with \emph{penalizedSVM}. The two-step ADMM-CD performs the best in terms of the model accuracy and model complexity. 

%\ccl{Is the above explanation in agree with the following result? Is two-step better than L1? }

\begin{table}[htbp] %the data here needs to be changed
	\begin{center}
		\caption{\label{realdata:svm}  Performances of ADMM and \emph{lpSolve} of sparse SVM on the microarray dataset.} 
		\scalebox{1}{
			\begin{tabular}{cccc}
				\vspace{1pt}\\
	\hline			
				& Train error & Test error & Size  \\
     \hline   
  $\ell_1$-ADMM-CD & 0.000 (0.00) & 0.000 (0.00) & 27.73 (0.25)  \\
  
  $\ell_1$-ADMM-prox  & 0.002 (0.00) & 0.003 (0.00) & 65.11 (2.06)   \\
  
 $\ell_1$-\emph{lpSolve}  & &  \ding{55}  &  \\
  
  Two-step ADMM-CD & 0.000 (0.00) &  0.000 (0.00) & 23.84 (0.31) \\
  
  Two-step ADMM-prox & 0.003 (0.00) & 0.003 (0.00) &  76.10 (1.08)  \\
 Two-step \emph{lpSolve}  & &  \ding{55}  & \\
 
                   \hline
			\end{tabular}
		}
	\end{center}
\end{table}

\begin{comment}

					     \hline
				   &  $\ell_1$-ADMM-CD & $\ell_1$-ADMM-prox & \emph{\emph{lpSolve}} \\
				   \hline
                Train Error & 0.01(0.00) & 0.00(0.00) &  0.04(0.02)  \\

                Test Error   & 0.02 (0.00) &  0.00(0.00)  &  0.48 (0.06) \\

                 CPU & 190.91(1.49)  & 218.6(17.67) & 314.14(1.03)  \\ 
                 Model size    & 3.00(0.02) & 177.28(1.61) & 21.49(0.33) \\
\end{comment}

\section{Conclusion}
We propose efficient and parallelizable algorithms for sparse SVM based on a 3-block ADMM with feature space split. The proposed algorithm is computationally advantageous in high and ultra-high dimensional settings. The numerical experiments suggest that the proposed method has comparable performances compared with standard solvers and maintains satisfactory performances in ultra-high dimensional settings. The computational framework can also be generalized to a wide range of statistical model fitting problems. We also demonstrate the application of the proposed algorithm in econometrics with a real data from a Chinese supermarket.

% what to prove
% General case the weighted L1 is convergent, then at first step, Lasso penalized the algorithm can converge to the unique value; that is beta0 is estimated to meet the requirement (write down which requirement)
% on the second stage, again the algorithm will converge to beta1, since again it is a convex problem.
% write down the proof for the three block case.
% the three-block is equivalent to a two-block case. 
% we can write down the proof for each of them actually. just follow the proof of the general case.
% because the optimality condition is different for each problem. Yeah, let's do that.


\newpage
%\bibliographystyle{abbrvnat}
%\bibliography{example_paper}

\bibliographystyle{elsarticle-num}

\bibliography{main}

\end{document}